%% file: paper.tex
\documentclass[preprint2]{emulateapj-rtx4}
\usepackage{graphicx,amsmath,url}
\usepackage{bm}
\graphicspath{{./fig/}{./png/}}
\input apj
\input macros

\newcommand{\HH}{\overline{h}}

\newcommand {\BBM}{\overline{\BB}}

\newcommand{\aaa}{\mbox{\boldmath $a$}{}}{}

\newcommand{\aO}{$\alpha\Omega$\ }
\newcommand{\aOO}{$\alpha\Omega$}
\newcommand{\aaO}{$\alpha^2\Omega$\ }
\newcommand{\alp}{$\alpha^2$\ }
\newcommand{\alpp}{$\alpha^2$}

\newcommand{\rms}{\text{rms}}

\begin{document}

\title{Catastrophic quenching in \aO dynamos revisited}
\author{Alexander Hubbard$^{1,2}$  \& Axel Brandenburg$^{1,3}$}
\affil{
$^1$ NORDITA, AlbaNova University Center, Roslagstullsbacken 23,
SE 10691 Stockholm, Sweden\\
$^2$ Max Planck Institut f\"ur Astronomie, K\"onigstuhl 17, D-69117
Heidelberg, Germany \\
$^3$Department of Astronomy, AlbaNova University Center,
Stockholm University, SE 10691 Stockholm, Sweden
}
\email{alex.i.hubbard@gmail.com
($ $Revision: 1.78 $ $)
}

\begin{abstract}

At large magnetic Reynolds numbers, magnetic helicity evolution plays an
important role in astrophysical large-scale dynamos.
The recognition of this fact led to the development of the dynamical $\alpha$ quenching formalism,
which predicts catastrophically low mean fields in open systems.
Here we show that in oscillatory \aO dynamos this formalism predicts an
unphysical magnetic helicity transfer between scales.
An alternative technique is proposed where this artifact is removed
by using the evolution equation for the magnetic helicity of the total
field in the shearing--advective gauge.
In the traditional dynamical $\alpha$ quenching formalism, this can be
described by an additional magnetic helicity flux of small-scale fields
that does not appear in homogeneous $\alpha^2$ dynamos.
In \aO dynamos, the alternative formalism is shown to lead to larger
saturation fields than what has been obtained in some earlier models
with the traditional formalism.
We have compared the predictions of the two formalisms to results of
direct numerical simulations, finding that the alternative formulation
provides a better fit.  This suggests that worries about catastrophic
dynamo behavior in the limit of large magnetic Reynolds number
are unfounded.
\end{abstract}


\section{Introduction}

While the possibility, and indeed need, for astrophysical dynamos was
recognized quite early \citep{Lar19},
the study of dynamos has since been troubled by a number of problems.
Cowling's anti-dynamo theorem \citep{Cowling} initially
appeared to demonstrate that the entire concept was impossible,
though \cite{Par55} eventually discovered the physics behind what
has come to be called the $\alpha$ effect.
Cowling's anti-dynamo theorem was finally shown to be largely inapplicable
by analytically solvable dynamos such as the Herzenberg
dynamo \citep{Herzenberg}.  Once the possibility of dynamo action was demonstrated, the development
of mean-field $\alpha$ dynamo theory followed \citep{Steenbeck},
which describes the generation of poloidal field
from toroidal fields.

While the generation of toroidal magnetic fields from
sheared poloidal fields is straightforward through the $\Omega$ effect,
the reverse process is tricky.  Without it however, dynamo action is impossible.
The $\alpha$ effect, which relies on helicity (twist) in the fluid motion,
allows for the generation of strong large-scale magnetic fields such
as those observed in the Universe.
It can drive dynamo action on its own (\alp systems),
but as shear is ubiquitous in astrophysics,
shear-amplified dynamo action is generally expected to outperform \alp dynamos.
Accordingly, \aO dynamos, which combine the effects,
are expected to be the dominant type of natural astrophysical dynamo
\citep{Shear}.

More recently however, there were indications,
first suggested by \cite{Vainshtein},
that the $\alpha$ effect decreases catastrophically
already for weak mean fields
in the limit of large magnetic Reynolds number
(i.e.\ low non-dimensionalized resistivities).
Such behavior would imply that mean-field dynamos driven by the
$\alpha$ effect could not generate the observed large-scale magnetic fields.
This claim stymied the field of large-scale dynamos for the 1990s.
While strong fields are observed in nature,
the theoretical understanding appeared to have been cut down.
Eventually it was recognized that this behavior is not generally applicable, being restricted
to two-dimensional systems, or to homogeneous (non-dynamo generated)
mean fields \citep{BB02},
and large-scale dynamo simulations became common \citep{B01,BD01}.
These new simulations occurred alongside the realization
that magnetic helicity conservation,
through the dynamical $\alpha$ quenching formalism,
provides an excellent theoretical understanding
of the saturation of $\alpha$--effect
dynamos:  the build-up of small-scale magnetic helicity quenches the
$\alpha$ effect \citep{FB02}.
Even so, the question of catastrophic
quenching has remained open, with indications of saturated large-scale field strength
decreasing with increasing
magnetic Reynolds number for shearing sheets and open \alp systems
\citep{BS05b}.  Further, while the saturation
field strength in \alp systems with periodic or perfectly conducting
boundaries has been found to be independent of the resistivity for adequately
(and in practice modestly)
super-critical $\Rm$, the timescale to reach
saturation increases linearly with $\Rm$ \citep{B01}.
This has led to the study of magnetic helicity fluxes
\citep{Vishniac, BS04, Mitra10, AdvGauge},
where the hope is that, because the build-up of small-scale magnetic helicity
quenches the $\alpha$ effect,
stronger and faster growing dynamos should be possible if the helicity is,
instead, exported (as it cannot be
destroyed except through the action of true, i.e.\ microphysical, dissipation).

Probing the reality of catastrophic quenching is naturally difficult.  Analytical theory is impossible,
and direct numerical simulations are limited to $\Rm$ that, while significantly super-critical
for many systems, are
nevertheless orders of magnitude below those of astrophysical systems.
The dynamical $\alpha$ quenching formalism allows
probing large $\Rm$ in systems it can handle,
but its validity there cannot, of course, be directly verified.
While the evidence for and against catastrophic quenching is limited, resolving
the issue is a crucial step in advancing dynamo theory.

The continued improvement in techniques
to measure turbulent dynamo coefficients from simulations
has enabled new approaches to evaluating different
formulations of the dynamical quenching formalism.
In particular, the test-field method \citep{Sch05,Sch07} has been used
to rule out the possibility of catastrophic quenching of the turbulent
magnetic diffusivity $\etat$ in $\alpha^2$ dynamos \citep{BRRS08}.
Recent advances in the theory of magnetic helicity fluxes in the presence
of shear \citep{Shear} have led us to
continue these developments in dynamical quenching by revisiting earlier
results from shearing systems.
Somewhat surprisingly, these developments return 
the 1-D $\alpha$ dependent models to the first
0-D $\alpha$ dependent models \citep{BB02}.
In addition, we shall extend here earlier
numerical studies of \alp dynamos in open systems.

\section{Mean-field modeling}
\label{MFM}

\subsection{Mean-field dynamo action}

We reproduce here some basic results of mean-field modeling.
The dynamos we will consider are in the family of \alpp, \aOO, and
\aaO dynamos, i.e.\ dynamos
where the conversion of toroidal field to poloidal field occurs through the $\alpha$ effect,
while the conversion of poloidal to toroidal field occurs through
the $\alpha$ effect, the $\Omega$ effect,
and a combination of the two.
In practice, because some conversion of
poloidal field to toroidal field through the $\alpha$ effect
is always present, \aO dynamos
are an approximation in the limit that the $\Omega$ effect is much stronger than the $\alpha$ effect.
All three dynamos, in an infinite, (shearing-) periodic system arise from the same
eigenvalue problem.

Although we will focus in this work on the discussion of results from
numerical simulations, 
these results are better understood in terms of linear theory.
We assume a standard, isotropic homogeneous $\alpha$,
turbulent resistivity $\etat$, and consider a system with shear
velocity $\UU_S=S x \ithat{\yy}$.
We make a standard mean-field decomposition using $xy$-planar averaging throughout,
with over-barred upper-case variables denoting averaged quantities and lower-cased non-overbarred referring to
fluctuating quantities, e.g.,
\begin{align}
&\BB=\meanBB+\bb, \\
&\meanBB \equiv \frac{1}{L_x L_y} \int_x \int_y \BB \,\dd x \,\dd y.
\end{align}
An important deviation from this notation is the magnetic helicity $h \equiv \AAA \cdot \BB$, where we use
\begin{align}
& \HH \equiv \frac{1}{L_x L_y} \int_x \int_y \AAA \cdot \BB \,\dd x \,\dd y, \\
& \HH_m \equiv \meanAA \cdot \meanBB, \\
& \HH_f \equiv \HH-\HH_m = \overline{\aaa \cdot \bb},
\end{align}
i.e., $\HH$ is the $xy$-averaged magnetic helicity density
and $\HH_m$ is the magnetic helicity density
carried by the large-scale fields.  Note that while $\HH_f$ is the magnetic helicity density carried by the small-scale
fields, it is still a mean quantity.
We will further define $\kf$ as the wavenumber of the
energy-carrying scale of the turbulence,
$k_1$ as the scale of the mean-fields,
and $\Beq$ as the equipartition magnetic energy while working in units
for which $\Beq=\urms$, the rms turbulent velocity.

With these definitions and averaging choices, as long as $\meanUU=0$, the mean field equations are written as
\begin{align}
&\meanEMF=\alpha \meanBB-\eta_t \meanJJ, \label{meanEMF} \\
& \parder{\meanAA}{t}=\meanEMF-\eta \meanJJ, \\
&\meanBB=\nab \times \meanAA,
\end{align}
where $\meanJJ=\nab\times\meanBB$ is the mean current density
in units for which the vacuum permeability is unity,
and $\meanEMF$ is the mean electromotive force,
which is here expressed in terms of isotropic
$\alpha$-effect and $\eta_t$ is the turbulent resistivity \citep{B01}.
Accordingly, the mean-field problem
for a one-dimensional \aaO dynamo with $k_x=k_y=0$ and $k_z=k_1$
(i.e., averaging over the $xy$-plane) reduces to the eigenvalue problem
\EQ
\lambda\ithat\BB=
\left(\begin{array}{ccc}
-\etaT k_1^2 & -\ii\alpha k_1 & 0 \\
\ii\alpha k_1+S & -\etaT k_1^2 & 0 \\
0 & 0 & -\etaT k_1^2 \end{array}\right) \ithat\BB, \label{MFE}
\EN
where $\etaT=\eta+\etat$ is the total, microphysical and turbulent, resistivity.
The growing mode has eigenvalue and eigenvector
\begin{align}
&\lambda=|\alpha k_1|\sqrt{1-\ii Q}-\etaT k_1^2, \\
&\meanBB=B_0\left(\sin k_1z, \sgn(\alpha k_1) (1+Q^2)^{1/4} \sin(k_1z+\phi), 0\right), \label{Bform}
\end{align}
where
\EQ
Q \equiv  \frac{S}{\alpha k_1}
\EN
is a measure of the relative shear and
\EQ
\phi=1/2 \arctan Q 
\EN
is the phase between $\meanB_x$ and $\meanB_y$.
The growth rate of the \aaO mode is 
\EQ
\text{Re}\,\lambda=|\alpha k_1| \sqrt{(1+(1 +Q^2)^{1/2})/2}-\etaT k_1^2.
\EN

From the above, we draw some significant conclusions true for both \aO and the more general \aaO fields:
\begin{align}
& | \HH_m |=| \meanAA \cdot \meanBB|= \left| \frac{(1+Q^2)^{1/4}}{k_1} B_0^2 \sin \phi \right|, \label{HmaaO} \\
&\meanBB^2=B_0^2 \left(\sin^2 k_1 z+(1+Q^2)^{1/2}\sin^2 [k_1 z+\phi] \right),
\end{align}
i.e., the magnetic helicity density and the current helicity density
$C_m \equiv \meanJJ \cdot \meanBB$
of the mean-field are spatially uniform, while the amplitude of the mean-field
is not spatially uniform if $S \neq 0$ ($Q \neq 0$ and $\phi \neq \pi/2$).

We next make the \aO
approximation, assuming that $|Q|\gg 1$.  We also consider only the case of $\alpha, k_1, S \ge 0$ to
simplify notation (the other cases are analogous).  This implies that
\begin{align}
&(1+Q^2)^{1/4} \simeq Q^{1/2} \gg 1, \\
& \phi= \frac{\pi}{4}, \\
& \text{Re}\,\lambda=|\alpha k_1| \sqrt{Q/2}-\etaT k_1^2.
\end{align}
At constant $\etaT$ and $S$ then, the system will be stationary for $\alpha=\alpha_c$ such that
\EQ
\alpha_c=\frac{2\etaT^2 k_1^3}{S}, \quad Q^{1/2}=\sqrt{\frac 12} \frac{S}{\etaT k_1^2}.
\EN
Further, in such a state we have
\EQ
\frac{\HH_m}{\bra{\meanBB^2}}=\frac{2\etaT k_1}{S}. \label{aOHoverB}
\EN
The $\alpha$ effect from maximally helical turbulence has
$\alpha \sim \etat \kf$, so
the mean magnetic field of an \aO dynamo is expected to have very low helicity.
As we will see, this is an important consideration.

\subsection{Catastrophic $\alpha$--quenching}

Given the level of interest, it should be noted that ``catastrophic''
$\alpha$--quenching has not been consistently defined.
We will choose the following definitions:
\begin{itemize}
\item{Type 1} catastrophic quenching is probably the most extreme case.  Here, the \emph{saturated
mean-field strength} varies inversely with $\Rm$ (or some non-negligible negative power or similar).
\item{Type 2} catastrophic quenching is well understood in an \alp dynamo in a triply-periodic setup
as discussed in \Sec{Sec:a2}.  Here,
the \emph{time} required for final saturation scales linearly with $\Rm$ (or some non-negligible positive
power thereof).
\end{itemize}
A well known example of Type 2 catastrophic quenching is seen in the simulations
of \cite{B01}, while Type 1 catastrophic quenching has been suspected to
occur in the simulations of \cite{BD01,BS05b}, but this will be challenged
by the present work.

It should be noted that both Type 1 and 2 quenchings might be less than fully catastrophic in practice.
A system which rapidly reaches an $\Rm$-independent
field strength and then resistively decays
could be Type 1 and yet have a significant field for all relevant times.
Similarly, a system could
take a prohibitive resistive time to fully saturate,
but already reach significant field strengths
on dynamical times.

Given the name $\alpha$--quenching, it would be appropriate to define a quenching type
based on the value of $\alpha$.
Such a definition is quite difficult however, as in the saturated regime
the dynamo-driving effect
counterbalances resistive decay, so the \emph{net} dynamo-driving terms,
including the turbulent resistivity that must accompany an $\alpha$-effect,
are expected to vary with $\eta$ (and so with $\Rm^{-1}$).

\subsection{Dynamical $\alpha$--quenching}
\label{Sec:a2}

Dynamical $\alpha$--quenching is a theoretical advance,
first introduced by \cite{KR82} and more recently seen in \cite{BB02},
that uses the magnetic $\alpha$-effect
of \cite{PFL}.
Under that hypothesis, the actual $\alpha$ effect in a system can be decomposed
into a component due to the kinetic effect, $\alpha_K$,
and a component due to the backreaction of the
magnetic fields on the flow, $\alpha_M$:
\EQ
\alpha=\alpha_K+\alpha_M, \quad
\alpha_K \simeq -\frac{\tau}{3}\overline{\oo \cdot \uu}, \quad
\alpha_M=\frac{\tau}{3\rho}\overline{\jj \cdot \bb}. \label{alphas}
\EN
The mean current helicity density of the small-scale field,
$\overline{\jj \cdot \bb}$, is not a tractable quantity, but in general it is well
approximated by the mean magnetic helicity density of the small-scale field,
$\HH_f=\overline{\aaa \cdot \bb}$, through $\overline{\jj\cdot\bb} \simeq k_f^2
\overline{\aaa \cdot \bb}$;
see \cite{Mitra10} for details and results in an inhomogeneous system.
Recall that under this definition, \emph{small-scale magnetic helicity}
is a \emph{mean} quantity.

The mean small-scale magnetic helicity can be found by subtracting the evolution equation of the
large-scale magnetic helicity from that of the total helicity.
This can be determined from the uncurled induction equation
[see Section 3 of \cite{BS05}, noting the sign
error for the $\nab \phi$ terms in their Equations (3.33) and (3.44)],
\begin{align}
& \EE=-\UU \times \BB +\eta \JJ, \\
&\parder{\AAA}{t}=-\EE-\nab \phi, \label{dAdt}\\
&\BB=\nab \times \AAA,
\end{align}
where $\EE$ is the electric field.  After some vector identities, we arrive at
\begin{align}
&\parder{\HH}{t}=-2 \eta \overline{\JJ \cdot \BB} - \nab \cdot \meanFFFF, \label{dhdt} \\
&\parder{\HH_m}{t}=2 \meanEMF \cdot \meanBB -2 \eta \meanJJ \cdot \meanBB -\nab \cdot \meanFFFF_m, \label{dhmdt}\\
&\parder{\HH_f}{t}=-2\meanEMF \cdot \meanBB-2\eta \overline{\jj \cdot \bb} -\nab \cdot \meanFFFF_f, \label{dhfdt}
\end{align}
where $\meanFFFF=\meanFFFF_m+\meanFFFF_f$ is the sum of
large-scale and small-scale magnetic helicity fluxes
and $\meanEMF \equiv \overline{\uu\times \bb}$; see Eq.~(\ref{meanEMF}).
Note that the contribution of $(\UU \times \BB) \cdot \BB=0$ in \Eq{dhdt} is split into finite terms of opposite sign $\pm \meanEMF \cdot
\meanBB$ in \Eqs{dhmdt}{dhfdt}.
The gauge term in \Eq{dAdt} is included in the flux terms;
for a complete discussion see \cite{Shear}.
Using $\overline{\jj \cdot \bb} \simeq k_f^2 \overline{\aaa \cdot \bb}$,
\Eq{dhfdt} can be evolved
in a mean-field simulation if a form for the flux term is assumed.  We call this \emph{traditional} dynamical
$\alpha$--quenching.  In homogeneous, periodic systems, such
as homogeneous \alp dynamos in triply periodic cubes, the flux term vanishes, and the concept behind
dynamical $\alpha$--quenching can be tested.  The application of dynamical $\alpha$--quenching to
this system predicts Type 2 quenching: there is an exponential growth phase
which ends when $\meanB^2/\Beq^2=k_1/\kf$ \citep{BB02}.
Subsequently, there is a resistively controlled saturation phase
with time $1/2 \eta k_1^2$, finally ending
at a saturated field strength of $\meanB^2/\Beq^2=\kf/k_1$ \citep{B01}.

Recent work suggests that the appropriate
\emph{ansatz} for the flux of mean small-scale magnetic helicity is diffusive, with sub-turbulent diffusion
coefficients \citep{HB10}.
However, recent work \citep{Shear} has also demonstrated that shear poses a unique
problem which can
be seen in the case of a shearing-periodic setup
at a moment when all quantities are periodic except for the 
imposed shear flow $\UU_S=Sx\ \ithat{\yy}$.
In that case, the helicity flux has a horizontal component,
$(\UU_S \times \BB) \times \AAA$, which is not periodic and has a finite divergence.
While the existence of this net flux through the
shearing-periodic boundaries might be unexpected,
the need for it can be simply explained.
The solution of an \aaO dynamo has spatially uniform large-scale helicity,
as quantified by \Eq{HmaaO},
but the $\meanEMF \cdot \meanBB$ term in \Eq{dhmdt} depends on $z$.  A flux term with a finite divergence is
required to balance the equation.
This flux term follows naturally from the requirement that whatever
terms the mean electromotive force produces in the evolution equation
for the magnetic helicity of the mean field, it should not affect the
evolution of magnetic helicity of the total field.
In other words, no terms involving $\meanEMF$ should appear in the
evolution equation for $\meanAA\cdot\meanBB+\overline{\aaa\cdot\bb}$.
Any term with $\meanEMF$ in  the equation for $\HH_m$ should thus be
absorbed by such a term with opposite sign in the equation for $\HH_f$.

To elucidate this further, let us consider the equation for
$\partial_t\meanAA=\meanEMF$ for the mean field.
Dotting this with $\meanBB$ gives the contribution $\meanEMF\cdot\meanBB$
for the production of $\meanAA\cdot\meanBB$.
We still need the contribution from $\meanAA\cdot\partial_t\meanBB$,
i.e., $\meanAA\cdot\nab\times\meanEMF$.
Using the identity
\EQ
\meanAA\cdot\nab\times\meanEMF=
\meanEMF\cdot\nab\times\meanAA+\nab\cdot(\meanEMF\times\meanAA),
\EN
we have
\EQ
\frac{\partial}{\partial t}\meanAA\cdot\meanBB
=2\meanEMF\cdot\meanBB+\nab\cdot(\meanEMF\times\meanAA)+...
\EN
where dots indicate the presence of other terms not involving
$\meanEMF$ for the full equation.
Thus, the evolution equation for $\overline{\aaa\cdot\bb}$
must then be of the form
\EQ
\frac{\partial}{\partial t}\overline{\aaa\cdot\bb}
=-2\meanEMF\cdot\meanBB-\nab\cdot(\meanEMF\times\meanAA)+...
\EN
so that the evolution of $\meanAA\cdot\meanBB+\overline{\aaa\cdot\bb}$
is not effected by the $\meanEMF$ terms.
In the traditional dynamical $\alpha$--quenching formalism, this
was only true of the $2\meanEMF\cdot\meanBB$ term, but the
divergence of $\meanEMF\times\meanAA$ had been ignored.

Allowing now for all the other terms in \Eqs{dhmdt}{dhfdt},
our full set of equations is
\begin{align}
&\parder{\HH_m}{t}=2 \meanEMF \cdot \meanBB -2 \eta \meanJJ \cdot \meanBB -
    \nab \cdot(\meanFFFF_m^{'}-\meanEMF\times\meanAA), \label{dhmdt2}\\
&\parder{\HH_f}{t}=-2\meanEMF \cdot \meanBB-2\eta \overline{\jj \cdot \bb} -
    \nab \cdot(\meanFFFF-\meanFFFF_m^{'}+\meanEMF\times\meanAA), \label{dhfdt2}
\end{align}
where $\meanFFFF_m^{'}$ is the resistive component of $\meanFFFF_m$.
When $\meanBB$ takes the form in \eq{Bform} and $\meanEMF=\alpha \meanBB-\etat\meanJJ$,
the $\nab \cdot (\meanEMF \times \meanAA)$ terms
cancel the $\meanEMF \cdot \meanBB$ terms.
\cite{BR00} have estimated that this $\meanEMF \times \meanAA$
flux can be important in the Sun.

If the flux term is not correctly handled, we can
expect the generation of artificial helicity ``hot-spots'' through the $\meanEMF \cdot \meanBB$ terms,
which will nonlinearly back-react on the dynamo through \Eq{alphas}.
While
an adequate diffusive flux may be able to smooth out such, this poses a clear potential difficulty
in applying dynamical $\alpha$--quenching to shearing systems.

If a mean-field model is solved in terms of the mean magnetic vector potential $\meanAA$ however, then $\HH_m$ is
known at every time step.
Thus, rather than evolving \Eq{dhfdt},
one can evolve \Eq{dhdt} to find $\HH_f=\HH-\HH_m$, avoiding
the $\meanEMF \cdot \meanBB$ terms.
One known difficulty with this alternate technique is that
spatially homogeneous components of $\meanAA$
may develop and cause spurious spatial variation in $\alpha_M$
when the latter is defined in terms of $\meanAA\cdot\meanBB$.
This homogeneous component arises from numerical noise:
as a constant $\meanAA$ is curl-free,
physically motivated equations cannot generate or, unfortunately, erase it.
Accordingly, we must artificially treat the issue by
subtracting out the volume averaged $\bra{\meanAA}_V$.
We refer to this technique of calculating $\HH_f$ as
\emph{alternate} dynamical $\alpha$--quenching.
For systems with no native spatial variations in $\alpha$, nor any instabilities
in the spatial variation of $\alpha$, this procedure will in practice return one
to the first attempts to apply dynamical $\alpha$--quenching using volume
averages \citep{BB02}.

\begin{figure}[t!]\begin{center}
\includegraphics[width=\columnwidth]{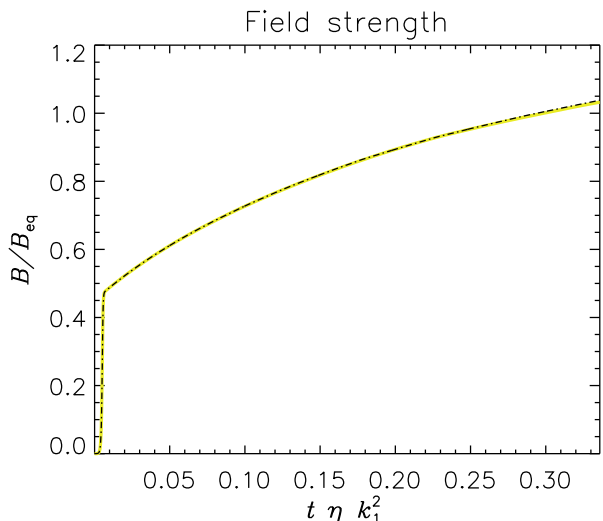}
\end{center}\caption{
Mean-field simulations for an \alp dynamo at $\Rm=10^3$,
comparing traditional (yellow/gray, solid/thick) and alternate
(black/dashed) dynamical quenching models in a system where they are formally identical.
\label{a2} }
\end{figure}

We use an \alp dynamo to test alternate dynamical $\alpha$--quenching against  traditional dynamical $\alpha$--quenching
(which, in this system, should be identical as there are no spatial variations and so no fluxes).  We show
the agreement in \Fig{a2}.  The small difference that develops is due to a smaller rms spatial noise of $\alpha_M$ in
the alternate quenching case.

\subsection{Investigation procedure}

Catastrophic $\alpha$--quenching lives in the high $\Rm$ regime, beyond the reach of current direct numerical simulation
or laboratory experiment.  This makes confirming or disproving its existence impossible.  The evidence for its existence
lies largely on mean-field simulations
\cite[see, e.g.,][]{BS05b,Gustavo10},
which confirm Type 2 quenching for homogeneous isotropic periodic \alp
dynamos.  Further, mean-field simulations using traditional $\alpha$--quenching have strongly suggested the
existence of Type 1 quenching for shearing systems.

While we cannot simulate \aaO dynamos at high $\Rm$, we are in a position
to run modest $\Rm$ shearing simulations
to compare the predictions of traditional quenching (with and without diffusive magnetic helicity fluxes) with those
of alternate quenching.
In the latter case, we do not include uncertain diffusive fluxes because
the magnetic helicity and therefore $\alpha$-effect are not expected to
exhibit spatial dependencies, which we confirm.

Our procedure then is to run a direct numerical simulation of an \aaO dynamo, extract the spatial dependency of
$\alpha$ and compare it with the results of mean-field theories.  Once mean-field theories have been weighed against
the evidence, we move to large $\Rm$ and examine the evidence for or against Type 1 and 2 quenching.

\section{Numerics}
\label{Numerics}

We perform mean-field numerical simulations for a shearing sheet, with
$\UU_S=S x \hat{\yy}$, and averaging performed over the $xy$ plane,
so mean quantities
are only a function of $z$, reducing the problem to a one-dimensional one.
Our mean field equations are evolved using the same algorithm as the {\sc Pencil Code}
(see below), but due to ease of implementation at the time, and low numerical load,
run using the Interactive Data Language (IDL).

We formulate the mean $\meanEMF$ through the standard formula
\eq{meanEMF},
where $\etat$ is assumed not to be quenched;
see \cite{BRRS08} for a numerical justification of this.
The total $\alpha$ is given by the sum of the kinetic $\alpha_K$, presumed constant,
and the magnetic $\alpha_M$.  Accordingly, $\partial \alpha/\partial t=\partial \alpha_M/\partial t$.
We solve the two systems of equations
\begin{eqnarray}
&&\parder{\alpha}{t}=-2\etat \kf^2 \left(\frac{\meanEMF \cdot \meanBB}{\Beq^2}+
     \frac{\alpha-\alpha_K}{\etat/\eta}\right) +\DDD_{\alpha} \nab^2 \alpha, \\
&&\parder{\meanBB}{t}=\nab \times \left(\meanEMF-\eta \meanJJ\right),
\end{eqnarray}
for traditional quenching, see, e.g., Equations (9.14) and (9.15) of \cite{BS05}, 
where the helicity fluxes have been cast in the form of diffusion terms
following the results of \citealt{HB10}, where it was found that the flux
was proportional to the gradient of the magnetic helicities.
The diffusive helicity flux
has diffusion coefficient $\DDD_{\alpha}$ which will be scaled to $\eta_t$.
Alternate quenching solves instead
\begin{eqnarray}
&&\parder{\HH}{t}=-2\eta\left(\meanJJ \cdot \meanBB +\alpha_M \Beq^2/\etat\right), \\
&& \alpha_M=\etat \kf^2 (\HH-\meanAA \cdot \meanBB)/\Beq^2, \\
&&\parder{\meanAA}{t}=\meanEMF-\eta \meanJJ.
\end{eqnarray}
Note that for alternate quenching we also enforce $\int_z \meanAA\,\dd z=0$ at
every timestep to avoid drifts in the magnetic vector potential.
The essential difference between the two approaches can be traced back to
mutually canceling contributions to the large-scale and small-scale
magnetic helicity flux of the form $\mp\meanEMF\times\meanAA$.

Our direct numerical simulations are made using the {\sc Pencil Code},
a finite-difference scheme sixth order in space and third order in time.
In the {\sc Pencil Code} runs, we use the test-field method (TFM)
to determine
components of the $\alpha$ tensor as a function of position.
For information on TFM, see \cite{BRS08} and \cite{RB10}.
 
\section{Measured $\alpha$ profiles}

\subsection{Direct simulation}
\label{directsimulation}

\begin{figure}[t!]\begin{center}
\includegraphics[width=\columnwidth]{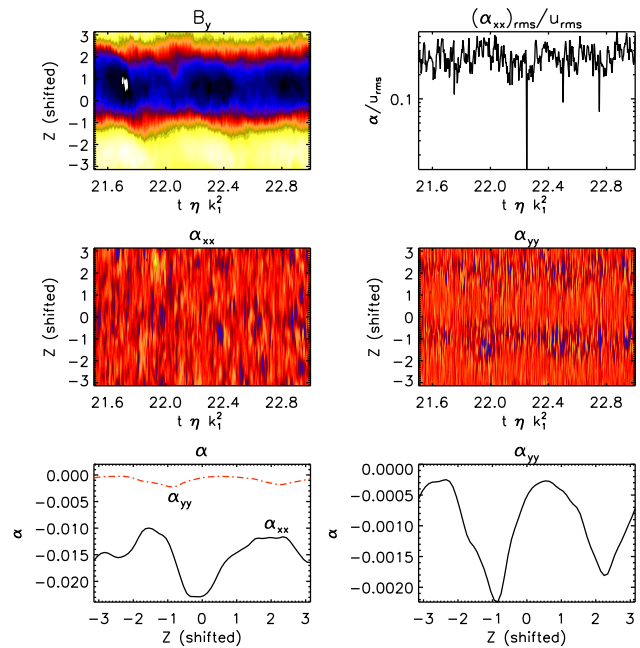}
\end{center}\caption{
Top left panel: $\meanBB_y$ in a frame comoving with the dynamo wave.
Top right panel: $\alpha_{xx}$.
Middle panels: butterfly diagrams of components of $\alpha(z,t)$.
Bottom panels: Time-averages of the middle panels.  Note differing y axis scales,
which implies that the quenching of $\alpha_{yy}$ is nearly uniform.  See \Sec{directsimulation}.
\label{AlpDNS} }
\end{figure}

In \Fig{AlpDNS} we present data for the $z$ dependence of $\alpha$ well into
the saturated regime for a direct simulation with $\Rm=27$ and $\kf/k_1=3$.
As in \cite{Shear}, the butterfly diagrams are shifted to the frame comoving 
with the traveling dynamo wave, as demonstrated in the top-left panel.  This allows
us to take meaningful time-averages while retaining spatial information.
In the top-right panel we show the volume \emph{rms} of $\alpha_{xx}$
in a semi-logarithmic plot, which demonstrates that the system
(including the small-scale fields in the TFM)
is in a steady state for the time interval considered.
The deep spike marks a reset of the test-fields \citep{Ossen02,Hetal09}.
The middle two contour plots show the two important components of $\alpha$ in the comoving frame.
The middle left panel shows $\alpha_{xx}$ which
aids the $\Omega$ effect in converting the poloidal $\meanBB_x$
into $\meanBB_y$, and the middle right
panel the vital $\alpha_{yy}$ which provides the conversion of the toroidal $\meanBB_y$ into $\meanBB_x$.

\begin{figure}[t!]\begin{center}
\includegraphics[width=\columnwidth]{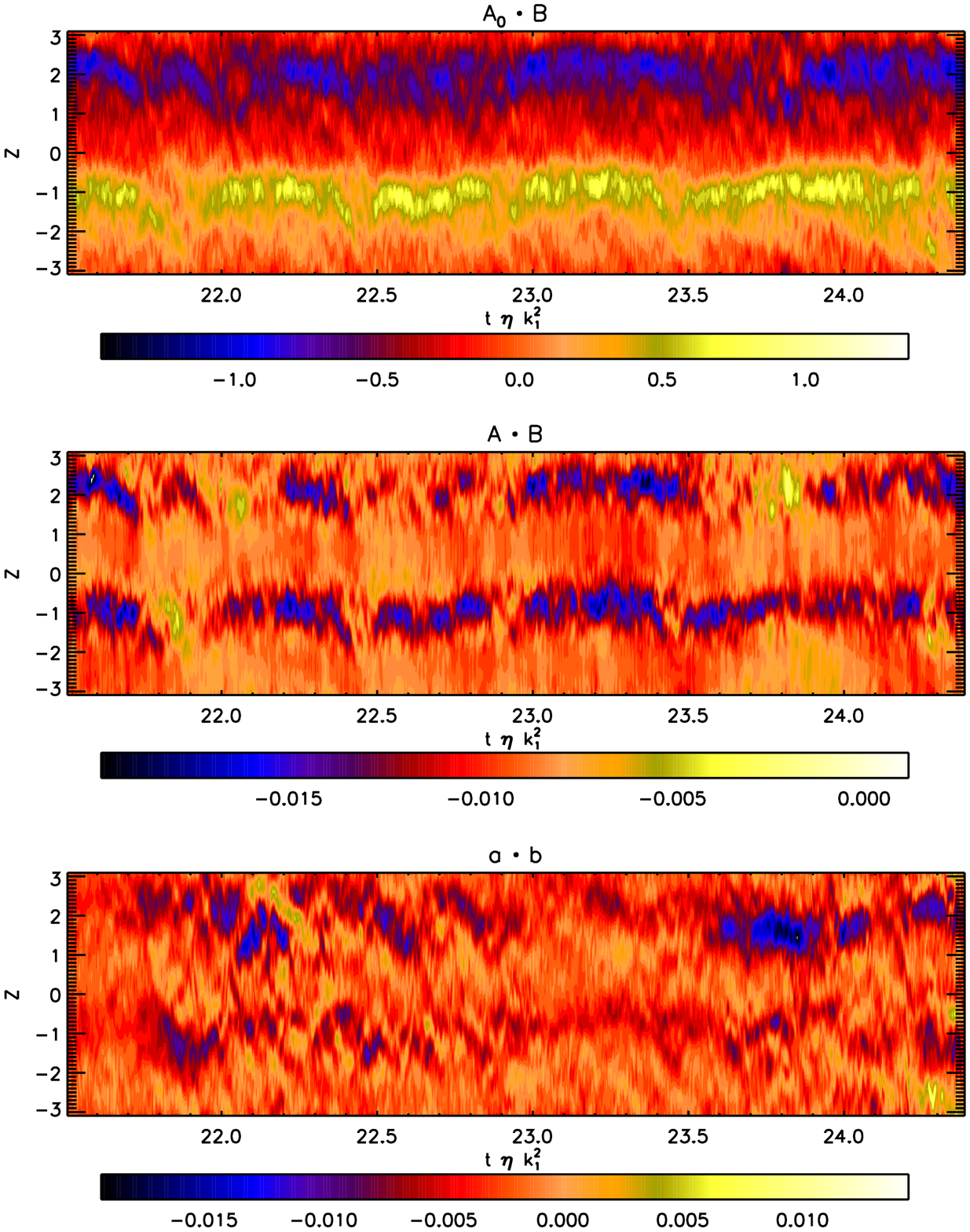}
\end{center}\caption{
Butterfly plots of magnetic helicity.
Top panel: $\bra{\AAA}_V \cdot \meanBB$,
i.e., the fictitious component of $\HH_m$
due to a spatially homogeneous component
of $\AAA$.  Middle panel: $\HH_m$, adjusted for the top panel.  Bottom panel:
$\HH_f$.  While there may be some spatial structure in the bottom panels,
it is intermittent in time, and the residual from a near-cancellation (the bottom two panels
use a very different scale than the top one, see the color bars).
\label{ab} }
\end{figure}

The bottom panels are time-averages of the middle panels.
It appears from the contour plot that $\alpha_{yy}$ shows
spatial variation, which is confirmed when a time average
(in the shifted domain) is taken as seen in the bottom right panel.  However, in the bottom left panel
it is clear that the actual result is that $\alpha_{yy}$ is strongly quenched compared to $\alpha_{xx}$.
This implies that the spatial variation
seen in $\alpha_{yy}$
is merely spatial variation in the residual $\alpha$ effect: the quenching itself is nearly uniform.

\Fig{ab}, for the same simulation, shows the difficulty mentioned in
\Sec{Sec:a2}, namely that a spatially homogenous component of
$\bra{\meanAA}_V$ can generate a spurious magnetic helicity signal.
We must also note here that the quenching is blatantly non-isotropic.
A study of this effect is beyond the scope of this paper: we expect
it to be a full project in its own right, and intend to study it as such.

\subsection{Mean-field approaches}

\begin{figure}[t!]\begin{center}
\includegraphics[width=\columnwidth]{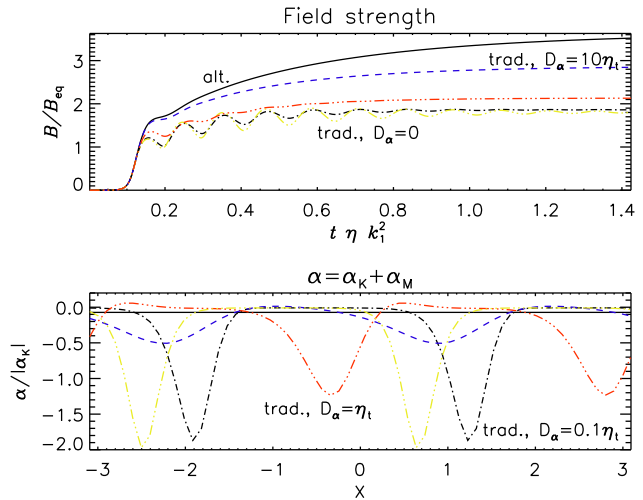}
\end{center}\caption{
Field and $\alpha=\alpha_K+\alpha_M$
(taken at the final time) profiles for mean-field simulations, approximately saturated.
The simulations are intended to be compared with that of \Fig{AlpDNS}, so
it has the same $\Rm=27$ and $\kf/k_1=3$.
Black/solid: alternate quenching formula.  Blue/dashed,
red, black and yellow/dash--dotted use the conventional quenching formula
with $\DDD_{\alpha}/\etat=10,1,0.1,0.01,0$, respectively.
\label{RunHlate} }
\end{figure}

In \Fig{RunHlate} we show energies and $\alpha_M$ profiles for mean-field simulations
similar to that of \Fig{AlpDNS} ($\Rm=27$, $\kf/k_1=3$).
The mean-field simulations use
traditional quenching with $\DDD_{\alpha}/\etat$ ranging from $0$ to $10$ and a run with
alternate quenching.
None of the traditional models match the uniform quenching that is measured
in \Fig{AlpDNS}, showing large spatial variability that derives from the $\meanEMF \cdot \meanBB$
term in \Eq{dhfdt}, not even the model with $\DDD_{\alpha}=10 \etat$.  The decrease in spatial variation
of $\alpha$ with increasing $\DDD_{\alpha}$ suggests that the traditional model could be made to
function with an adequate diffusion term, but this term would need to be absurd in scale (and would
hopelessly distort any simulation with ``real'' spatial variation in $\alpha_M$ that needs to be
correctly captured).  The alternate quenching formalism does result in the uniform quenching,
which is unsurprising as it eliminates the spatial forcing from $\meanEMF \cdot \meanBB$.

We take this as strong evidence that the alternate quenching formalism is superior to
traditional quenching
in sheared systems where drifts in $\meanAA$ are tractable
 -- and that results obtained with traditional quenching in the presence
of shear should be viewed with suspicion.

\section{Mean Field: Large magnetic Reynolds numbers}

\subsection{Early times}
\label{earlytimes}

\begin{figure}[t!]\begin{center}
\includegraphics[width=\columnwidth]{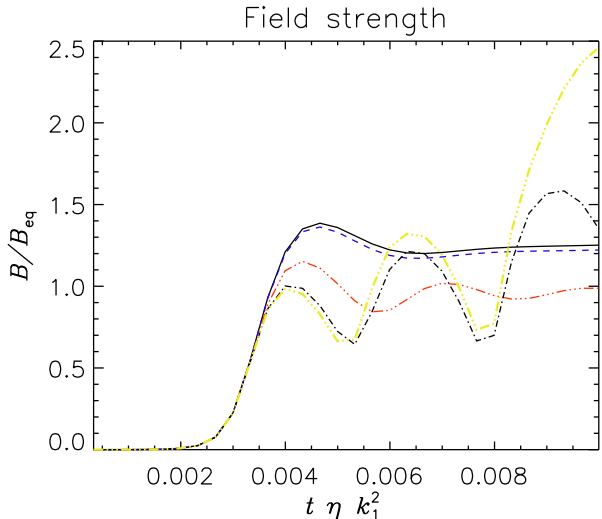}
\end{center}\caption{
Field magnitudes for mean-field simulations discussed in \Sec{earlytimes},
into the non-kinematic regime.  The simulations use $\Rm=10^3$, $\kf/k_1=3$.
Black/solid: alternate quenching formula.  Blue/dashed,
red, black and yellow/dash--dotted use the conventional quenching formula
with $\DDD_{\alpha}/\etat=10,1,0.1,0.01,0$, respectively.
Note the oscillations, which persist at some level even with $\DDD_{\alpha}=0.1\etat$.
\label{short} }
\end{figure}

For early times, the predictions of both dynamical $\alpha$--quenching formalisms
predict behavior similar to that of \alp dynamos: exponential growth of the mean fields
(and corresponding growth of $\alpha_M$) until the total $\alpha$ effect is reduced
enough that the growth rate is reduced to a fraction of its original self.
This occurs when $|\alpha|=|2\etaT^2 k_1^3/S|$, i.e., when
\EQ
\left|\frac{\tau}{3} \bra{\jj\cdot\bb}\right|=|\alpha_K|-|2\etaT^2 k_1^3/S|.
\EN
In terms of magnetic helicity, this becomes
\EQ
|\bra{\aaa \cdot \bb}| = \kf^{-2} \frac {3}{\tau} \left(|\alpha_K|-|2\etaT^2 k_1^3/S|\right).
\label{abright}
\EN
Using the standard approximations for fully helical turbulence
\citep{Sur_etal08},
namely $\tau \simeq 1/u_{\rms} k_f$, $\alpha_K \simeq u_{\rms}/3$ and
$\etat \simeq \tau u_{\rms}^2/3$, and writing $\Beq=\urms$, this reduces to
\EQ
|\bra{\aaa \cdot \bb}| \simeq
\left(1-\frac{2 k_1^3 \Beq}{3\kf^2|S|}\right) \frac{\Beq^2}{\kf}.
\EN
As the growth is rapid, we will have $\HH_m \simeq -\HH_f $ during this stage, and so
\EQ
|\HH_m| = \left(1-\frac{2 k_1^3 \Beq}{3\kf^2|S|}\right) \frac{\Beq^2}{\kf}.
\label{earlyhm}
\EN
However, \aO dynamo mean-fields are only weakly helical,
i.e.\ $\HH_m \ll \meanBB^2/k_1$.
Under the assumptions that the mean field is approximately stationary, and that the shear is strong
enough to use \Eq{aOHoverB} as an approximation, \Eq{earlyhm} implies that:
\EQ
\bra{\meanBB^2}=\frac{|S|}{2\etaT k}\HH_m
=\left(\left|\frac{S}{2\alpha_K k_1}\right|-\frac{k_1^2}{\kf^2}\right) \Beq^2.
\label{Ekin}
\EN
As we have made the \aO approximation that $|S|\gg|\alpha_K k_1|$,
this implies that an \aO field first feels
nonlinear effects for mean-field energies that are already
in super-equipartition.

In \Fig{short} we show the early evolution of a mean-field dynamo with
$\Rm=10^3$, $\alpha_K=-1/3$, $S=1$ and $\kf/k_1=3$.
\EEq{Ekin} implies that the exit from exponential growth occurs for $\meanB \approx 1.18 \Beq$,
which is well captured by alternate quenching, and traditional quenching with strong diffusive fluxes.

\subsection{Late times}

\begin{figure}[t!]\begin{center}
\includegraphics[width=\columnwidth]{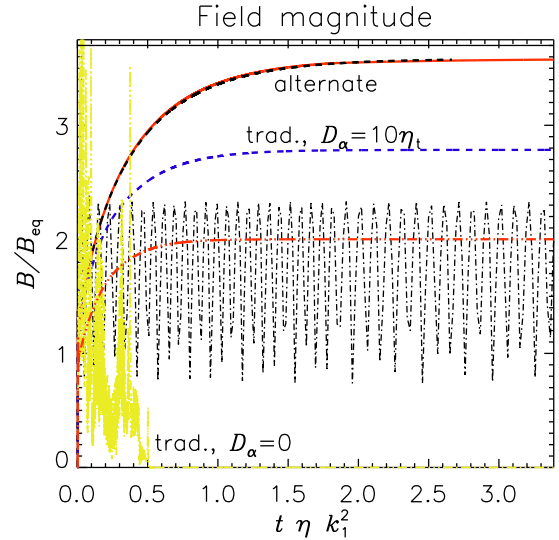}
\end{center}\caption{
Field magnitudes, for the same mean-field simulations as \Fig{short} ($\Rm=10^3$,
$\kf/k_1=3$), into the non-kinematic regime.
Red/solid and over-plotted black/dashed: alternate quenching formula for $\Rm=1000$ and $2000$
respectively.
Blue/dashed, red/dash-double-dotted, black/dash-dotted and yellow/dash-double-dotted
use the conventional quenching formula
with $\DDD_{\alpha}/\etat=10,1,0.1,0$, respectively.
The curve for $\DDD_{\alpha}/\etat=0.1$ is extremely strongly coarse-grained for visibility:
the oscillations (for that run) are in fact far more frequent than shown and would be a solid band
if plotted fully.
\label{long} }
\end{figure}

We can analytically estimate the final field strength of the dynamo for the
alternate quenching formalism,
while for traditional models the problem is nonlinear as can be seen in \Fig{RunHlate}.
The final state is achieved when $\partial \HH/\partial t=0$, i.e., when
$\overline{\JJ \cdot \BB}=\bra{\jj \cdot \bb}$.  Combining this with \Eq{aOHoverB} and
assuming that the shear is strong enough that $\alpha$ must be fully quenched,
$\alpha =\alpha_K+\alpha_M \simeq 0$, we find
\EQ
\BBM^2 \simeq |S/2\alphaK k_1|\,(\kf/k_1)^2\Beq^2. \label{latetimeB}
\EN
In \Fig{long} we show the late time evolution of the same mean field dynamos
as in \Sec{earlytimes}.  It is clear
that, without significant ($\DDD_{\alpha} > 0.1\etat$) helicity diffusion, the solution
for traditional quenching is unstable and drops to resistively small values.
This is not surprising as the problem becomes highly nonlinear.
However, with moderate diffusion the field strength behaves smoothly, with the final
energy level increasing with diffusion coefficient.  Even so, the saturation level of the
traditional quenching model with $\DDD_{\alpha}=10\etat$ is significantly below
that of the alternate quenching model.  While the diffusion does smooth out the helicity
hot-spots, the spatial fluctuations of $\alpha_M$ in \Fig{RunHlate} have a noteworthy
impact on the final dynamo state.  Finally, the saturation level of the alternate model
matches the estimate from \Eq{latetimeB} of $\BBM \simeq 3.7 \Beq$,
to within the limit that an adequate residual $\alpha$ is needed to sustain
the field against turbulent resistive decay.  Further, the overplotted black/dashed
alternate-quenching curve is for $\Rm=2000$, double that of the red/solid
alternate-quenching curve.  The overlay implies that we have reached an
asymptotic state independent of $\Rm$, which is in agreement with
earlier work assuming perfect spatial homogeneity \citep{BB02} and with
simulations \citep[see Figure~6 of][]{KB09}.

\section{Direct simulations of open systems}
\label{open}

\begin{figure}[t!]\begin{center}
\includegraphics[width=\columnwidth]{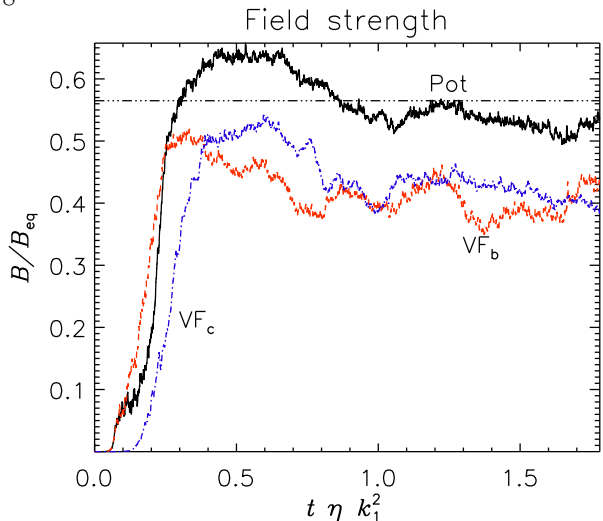}
\end{center}\caption{
Time series for \alp dynamos in open systems.  Black/solid: potential field extrapolation in the vertical direction.
Red/blue/dashed: vertical field condition on the vertical direction.  Potential field and VF$_b$ have $\Rm=156$
while VF$_c$ has $\Rm=86$.  The dash-double-dotted line corresponds to $1/k_f^{1/2}$, i.e. the energy
level associated with the end of the kinematic phase.
\label{VF} }
\end{figure}

Numerical resources limit our ability to probe the high $\Rm$ regime.  However, we have run
three simulations of \alp dynamos in an open system,
i.e., a system which can export magnetic helicity.
This system is the same one as considered in \cite{BS05b}: a helically forced cube, periodic in the
horizontal directions and with vertical field conditions in the vertical directions, which we have run for
$\Rm=86$ and $156$.
Additionally, as the vertical field condition is frequently used
instead of a proper vacuum
condition, we also performed a $\Rm=156$ run with potential field condition in the vertical directions.
The resulting time series are given in \Fig{VF}.  Our resolution was $128^3$, for runs
with $u_{\text{max}} \simeq 0.15$, $\urms \simeq 0.05$ and $\eta=2 \times 10^{-4}$
(for $\Rm=86$) or $\eta=10^{-4}$ (for the other two).
The velocity boundary has a stress-free
vertical condition, and the entropy a symmetric one.

Unlike the results reported in \cite{BS05b}, there is no clear indication
of a reduction in the strength of the mean field
for higher magnetic Reynolds number,
even though the runs were followed for resistive times.
However, the use of vertical field conditions as a proxy for vacuum
conditions appears to be a poor one.
Note that there does not appear to be a slow resistive phase.  This lack is expected
as the open boundaries allow the system to export total magnetic helicity (not just
helicity of the small-scaled field).
Thus, the system should reach a steady state where
exchanges of helicity through the boundary balance preferential destruction
of small-scale helicity on dynamical times, and for small total helicities.

\section{Discussion and conclusions}

We have used the test-field method to examine the predictions
of catastrophic $\alpha$--quenching 
resulting from dynamically--quenched mean-field models in shearing systems.
Formulations for dynamical $\alpha$--quenching which are superior for the problem of shearing systems
do not predict Type 1 catastrophic quenching (reduced field strength) but do predict Type 2 quenching
(long final saturation times),
extending results that do not allow for spatial variations of $\alpha$ \citep{BB02}
to models that do.
We have further revisited simulations of \alp dynamos in
open systems and, at admittedly quite modest $\Rm$, found no evidence of field strength scaling inversely 
with $\Rm$.

The picture we see now for $\alpha$--effect dynamos,
motivated by the concepts and formalism
of dynamical $\alpha$--quenching,
is one of exponential growth during a rapid initial saturation phase.
This phase ends when the magnetic
helicity in the small-scale fields is comparable to the helicity in the forcing that generates the $\alpha$--effect.
At this point, the total magnetic helicity in the system has not changed from its initial value.  If the system
is open, exchanges with the exterior (\Sec{open})
will tend to keep the \emph{total} magnetic helicity
roughly constant, and the system will then not evolve resistively.  On the other hand, if the system is closed
the preferential resistive destruction of magnetic helicity
of the small-scale field allows a further resistive growth phase.

It is important to note that the energy in the large-scale field is
bounded below by its helicity.
Weakly helical large-scale fields are possible, which can have
super-equipartition fields even at the end of the kinematic growth phase.
Weakly helical large-scale fields are a natural product of sheared system,
so rapid growth to sub-equi-, equi- and super-equipartition fields are all expected to occur in nature,
although all equi- and super-equipartition fields in the high $\Rm$ systems of astrophysics are
expected to be weakly helical.

\acknowledgements

This work was supported in part by the European Research Council under
the AstroDyn Research Project 227952.  Alexander Hubbard acknowledges
the additional support of a fellowship from the Alexander von Humboldt Foundation.
The computations have been carried out at the
National Supercomputer Centre in Link\"oping and the Center for
Parallel Computers at the Royal Institute of Technology in Sweden.


\end{document}

%% file: apj.tex
\newcommand{\EQ}{\begin{equation}}
\newcommand{\EN}{\end{equation}}
\newcommand{\EQA}{\begin{eqnarray}}
\newcommand{\ENA}{\end{eqnarray}}
\newcommand{\eq}[1]{(\ref{#1})}
\newcommand{\EEq}[1]{Equation~(\ref{#1})}
\newcommand{\Eq}[1]{Equation~(\ref{#1})}
\newcommand{\Eqs}[2]{Equations~(\ref{#1}) and~(\ref{#2})}

\newcommand{\Sec}[1]{Section~\ref{#1}}

\newcommand{\Fig}[1]{Figure~\ref{#1}}

\newcommand{\bra}[1]{\langle #1\rangle}

{}
{}
\newcommand{\meanFFFF}{\overline{\mbox{\boldmath ${\cal F}$}}{}}{}

{}
{}
\newcommand{\meanEMF}{\overline{\mbox{\boldmath ${\cal E}$}}{}}{}
{}
{}
{}
{}
{}
\newcommand{\meanAA}{\overline{\mbox{\boldmath $A$}}{}}{}
\newcommand{\meanBB}{\overline{\mbox{\boldmath $B$}}{}}{}
{}
{}
{}
{}
{}
{}
{}
{}
\newcommand{\meanJJ}{\overline{\mbox{\boldmath $J$}}{}}{}
{}
\newcommand{\meanUU}{\overline{\bm{U}}}

{}
{}
{}

\newcommand{\meanB}{\overline{B}}

{}

{}
{}

\newcommand{\alphaK}{\alpha_{\it K}}

\newcommand{\yy}{\mbox{\boldmath $y$} {}}

\newcommand{\uu}{\mbox{\boldmath $u$} {}}
\newcommand{\UU}{\mbox{\boldmath $U$} {}}

\def\bb{\bm{b}}
\newcommand{\BB}{\mbox{\boldmath $B$} {}}
\newcommand{\EE}{\mbox{\boldmath $E$} {}}
\newcommand{\jj}{\mbox{\boldmath $j$} {}}
\newcommand{\JJ}{\mbox{\boldmath $J$} {}}

\newcommand{\AAA}{\mbox{\boldmath $A$} {}}

\newcommand{\nab}{\mbox{\boldmath $\nabla$} {}}

\newcommand{\oo}{\mbox{\boldmath $\omega$} {}}

\newcommand{\ii}{{\rm i}}

\newcommand{\sgn}{{\rm sgn}  \, {}}

\newcommand{\DDD}{{\cal D} {}}
\newcommand{\dd}{{\rm d} {}}

\def\Rm{\mbox{\rm Re}_M}

\def\kf{k_{\it f}}

\def\urms{u_{\rm rms}}

\def\etat{\eta_{\it t}}

\def\etaT{\eta_{\rm T}}

\def\Beq{B_{\rm eq}}
\newcommand{\yjgr}[3]{ #1, {J.\ Geophys.\ Res.,} {#2}, #3}

\newcommand{\yapj}[3]{ #1, {ApJ,} {#2}, #3}

\newcommand{\yapjl}[3]{ #1, {ApJ,} {#2}, #3}

\newcommand{\yan}[3]{ #1, {Astron.\ Nachr.,} {#2}, #3}

\newcommand{\yana}[3]{ #1, {A\&A,} {#2}, #3}

\newcommand{\ygafd}[3]{ #1, {Geophys.\ Astrophys.\ Fluid Dyn.,} {#2}, #3}

\newcommand{\ymn}[3]{ #1, {MNRAS,} {#2}, #3}

\newcommand{\yjour}[4]{ #1, {#2}, {#3}, #4}
\newcommand{\yjourS}[3]{ #1, {#2}, #3}

%% file: macros.tex
\newcommand{\itover}[2]{\,\hspace{.0mm}#1{\!\hspace{-.0mm}#2}}

\newcommand{\ithat}[1]{\itover{\hat}{#1}}

\newcommand{\deriv}[3]{\frac{#3\hspace*{-.06em} {#1}}{#3\hspace*{.06em} {#2}}}

\newcommand{\parder}[2]{\deriv{#1}{#2}{\partial}}

%% file: paper.bbl
\begin{thebibliography}

\bibitem[Berger \& Ruzmaikin(2000)]{BR00}
Berger, M. A., \& Ruzmaikin, A.\yjgr{2000}{105}{10481}

\bibitem[Blackman 
\& Brandenburg(2002)]{BB02} Blackman, E.~G., \& Brandenburg, A.\ 2002, \apj, 579, 359

\bibitem[Brandenburg(2001)]{B01}
Brandenburg, A.\yapj{2001}{550}{824}

\bibitem[Brandenburg \& Dobler(2001)]{BD01}
Brandenburg, A., \& Dobler, W.\yana{2001}{369}{329}

\bibitem[Brandenburg \& Sandin(2004)]{BS04}
Brandenburg, A., \& Sandin, C.\yana{2004}{427}{13}

\bibitem[Brandenburg \& Subramanian(2005a)]{BS05}
Brandenburg, A., \& Subramanian, K.\yjour{2005a}{Phys.\ Rep.}{417}{1}

\bibitem[Brandenburg \& Subramanian(2005b)]{BS05b} Brandenburg, A., 
\& Subramanian, K.\ 2005b, Astron.\ Nachr., 326, 400 

\bibitem[Brandenburg et al.(2008a)]{BRRS08} Brandenburg, A., 
R\"adler, K.-H., Rheinhardt, M., \& Subramanian, K.\yapjl{2008a}{687}{L49}

\bibitem[Brandenburg et al.(2008b)]{BRS08} Brandenburg, A., 
R\"adler, K.-H., \& Schrinner, M.\yana{2008b}{482}{739}

\bibitem[Candelaresi et al.(2011)]{AdvGauge} Candelaresi, S., 
Hubbard, A., Brandenburg, A., 
\& Mitra, D.\ 2011, Physics of Plasmas, 18, 012903 

\bibitem[Cowling(1933)]{Cowling} Cowling, T.~G.\ 1933, \mnras, 
94, 39 

\bibitem[Field 
\& Blackman(2002)]{FB02} Field, G.~B., \& Blackman, E.~G.\ 2002, \apj, 572, 685 

\bibitem[Guerrero et al.(2010)]{Gustavo10} Guerrero, G., 
Chatterjee, P., \& Brandenburg, A.\ 2010, \mnras, 409, 1619 

\bibitem[Herzenberg(1958)]{Herzenberg} Herzenberg, A.\ 1958,
Phil. Trans. R.  Soc. A, 250, 543 

\bibitem[Hubbard \& Brandenburg(2010)]{HB10}
Hubbard, A., \& Brandenburg, A.\ygafd{2010}{104}{577}

\bibitem[Hubbard \& Brandenburg(2011)]{Shear}
Hubbard, A., \& Brandenburg, A.\ 2011, \apj, 727, 11 

\bibitem[Hubbard et al.(2009)]{Hetal09}
Hubbard, A., Del Sordo, F., K\"apyl\"a, P. J., \& Brandenburg, A.\ymn{2009}{398}{1891}

\bibitem[K\"apyl\"a \& Brandenburg(2009)]{KB09}
K\"apyl\"a, P. J., \& Brandenburg, A.\yapj{2009}{699}{1059}

\bibitem[Kleeorin \& Ruzmaikin(1982)]{KR82}
Kleeorin, N. I., \& Ruzmaikin, A. A.\yjour{1982}{Magne\-to\-hydro\-dynamics}{18}{116}

\bibitem[Larmor(1919)]{Lar19}
Larmor, J.\yjourS{1919}{Rep. Brit. Assoc. Adv. Sci.}{159}

\bibitem[Mitra et al.(2010)]{Mitra10} Mitra, D., Candelaresi, 
S., Chatterjee, P., Tavakol, R., 
\& Brandenburg, A.\ 2010, Astron.\ Nachr., 331, 130 

\bibitem[Ossendrijver et al.(2002)]{Ossen02}
Ossendrijver, M., Stix, M., Brandenburg, A., \&
R\"udiger, G.\yana{2002}{394}{735}

\bibitem[Parker(1955)]{Par55}
Parker, E. N.\yapj{1955}{122}{293}

\bibitem[Pouquet et al.(1976)]{PFL} Pouquet, A., Frisch, 
U., \& Leorat, J.\ 1976, Journal of Fluid Mechanics, 77, 321 

\bibitem[Rheinhardt \& Brandenburg(2010)]{RB10}
Rheinhardt, M., \& Brandenburg, A.\ 2010, \aa, 520, A28

\bibitem[Schrinner et al.(2005)]{Sch05} Schrinner, M., R\"adler, K.-H.,
Schmitt, D., Rheinhardt, M., Christensen, U.\yan{2005}{326}{245}

\bibitem[Schrinner et al.(2007)]{Sch07} Schrinner, M., R\"adler, K.-H.,
Schmitt, D., Rheinhardt, M., Christensen, U. R.\ygafd{2007}{101}{81}

\bibitem[Steenbeck et al.(1966)]{Steenbeck} Steenbeck, M., Krause, F.,  
R\"adler, K.-H.\ 1966, Zeitschr.\ Naturforsch.\ A, 21, 369 

\bibitem[Sur et al.(2008)]{Sur_etal08}
Sur, S., Brandenburg, A., \& Subramanian, K.\ymn{2008}{385}{L15}

\bibitem[Vainshtein 
\& Cattaneo(1992)]{Vainshtein} Vainshtein, S.~I., \& Cattaneo, F.\ 1992, \apj, 393, 165 

\bibitem[Vishniac 
\& Cho(2001)]{Vishniac} Vishniac, E.~T., \& Cho, J.\ 2001, \apj, 550, 752 

\end{thebibliography}
